\documentstyle[12pt]{article}

\begin{document}

\vskip1truecm

{\flushright IC/IR/99/7~~~~~~~~~~~~~~~~~~~~~~~~~~~~~~~~~~~~~~~~~~~~~~~~~~~~~~~~~
~~~~~~~~~ LPMO/02/99}
\vskip2truecm
\begin{centering}
{\large\bf An Example of $Z_{N}$--Graded Noncommutative Differential
Calculus} 

\vskip1truecm
{\bf A.E.F. Djemai} \\
\small Abdus Salam International Centre for Theoretical Physics, Trieste 34100, 
Italy\\
\small and\\
\small 
Institut de Physique,
Universit\'e d'Oran Es--s\'enia, 31100, Oran,
Algeria.\\ 
\vskip0.5truecm
and\\
\vskip0.5truecm
{\bf H. Smail}\\
\small Institut de Physique, Universit\'e 
d'Oran Es--s\'enia, 31100, Oran, Algeria\\

\vskip0.5truecm
July 1999
\vskip2truecm
\begin{abstract}
In this work, we consider the algebra $M_{N}(C)$ of $N\times N$ matrices
as a cyclic quantum plane. We also analyze the coaction of the quantum group
${\cal F}$ and the action of its dual quantum algebra ${\cal H}$ on it.
Then, we study the decomposition of $M_{N}(C)$ in terms of the quantum
algebra representations. Finally, we develop the differential algebra of the
cyclic group $Z_{N}$ with $d^{N} = 0$, and treat the particular case $N = 3$.
\end{abstract}
\end{centering}
\newpage

\section{Introduction}

In the last decade, the concept of the noncommutative differential geometry,
{\cite 1} has been extensively developed. The most simple example of
noncommutative differential geometry based on derivations is given by the
Grassmannian of the matrix algebra ${\cal M}_{N} = M_{N}(C)$,{\cite 2}.
The matrix algebra ${\cal M}_{N}$ can also be considered as a {\bf cyclic quantum
plane} ($q^{N} = 1$) on which a coaction of quantum group ${\cal F}$
and an action of its dual ${\cal H}$ are naturaly defined, and the associated
Wess--Zumino differential complex is constructed, ({\cite 3} and references
therein) .
Moreover, the notion of graded $q$--differential algebra with the condition
$d^{N} = 0$, has been recently introduced,{\cite 4}.\\
The main aim of this work is to study the noncommutative differential
geometry of the cyclic group $Z_{N}$, viewed as the subalgebra
${\cal M}_{N}^{diag}$ of diagonal matrices of ${\cal M}_{N}$, as an example
of $Z_{N}$--graded noncommutative differential calculus.\\
This work is organized as follows : In section 2, we give a presentation of the space
$M_{N}(C)$ as a cyclic Manin plane. In sections 3, we present the coaction
and the action of the quantum group ${\cal F}$ and its dual ${\cal H}$ on
${\cal M}_{N}$ respectively and study the reduction of
${\cal M}_{N}$ under the representation of ${\cal H}$. In section 4, we
construct the noncommutative differential complex of the cyclic group $Z_{N}$
with a $Z_{N}$--graded differential $d$, i.e. $d^{N} = 0$.
Finally, in section 5 we treat in details the case $N = 3$. The section 6 is
devoted to some conclusions and perspectives.

\section{${\cal M}_{N} \equiv M_{N}(C)$ as a cyclic quantum plane}
The algebra of $N \times N$ matrices can be generated by two elements $x$ and $y$
obeying the relations :
\begin{equation}
x y = q y x  \label{1} 
\end{equation}
\begin{equation}
x^{N} = y^{N} = {\bf 1}   \label{2}
\end{equation}
where $q$ denotes a N--th root of unity :
\begin{equation}
q^{N} = 1~~ ,~~q \ne 1~~,~~ \sum_{n}^{N-1} q^{n} = 0~~,~~ q^{n} = q^{n-N}~~,
~~n \in Z          \label{3}
\end{equation}
and ${\bf 1}$ is the $N \times N$ unit matrix. \\
Explicitly $x$ and $y$ can be represented by the matrices:
\begin{eqnarray*}
x = \left( \begin{array}{ccccccc}
       0&1&.&.&...&.&. \\
       0&0&1&.&...&.&. \\ 
       .&.&0&1&...&.&. \\
       .&.&.&.&...&.&. \\
       .&.&.&.&...&.&. \\
       0&.&.&.&...&0&1 \\
       1&0&.&.&...&.&0 
       \end{array}
       \right)~~~,~~~
y = \left( \begin{array}{ccccccc}       
       1 & . & .   & . & . & . & . \\ 
       . & q & .   & . & . & . & .  \\
       . & . &q^{2}& . & . & . & .  \\
       . & . & .   & . & . & . & .  \\ 
       . & . & .   & . & . & . & .  \\
       . & . & .   & . & . & . & .  \\
       . & . & .   & . & . & . & q^{N-1}
       \end{array}
       \right) .
\end{eqnarray*}
We call the algebra generated by elements $x$ and $y$ satisfying the relations
(\ref{1}) and (\ref{2}) the {\bf cyclic quantum plane} ${\cal M}_{N} \equiv
M_{N}(C)$. As a $N^{2}$--dimensional vector space, ${\cal M}_{N}$ is spanned
by the following basis:
\begin{eqnarray*}
\{\alpha^{rs} = x^{r} y^{s} ; r,s = 0,1,2,.......,N-1\}
\end{eqnarray*}
and is endowed by the following internal law :
\begin{eqnarray*}
\alpha^{rs}.\alpha^{mn} = f^{(rs)(mn)}~_{(kl)} \alpha^{kl}
\end{eqnarray*}
where $x^{r}y^{s} = q^{sr}y^{s}x^{r}$ and :
\begin{eqnarray*}
f^{(rs)(mn)}~_{(kl)} = q^{-ms} \delta^{r+m}~_{k} \delta^{s+n}~_{l}.
\end{eqnarray*}
The {\bf noncommutativity} of the elements of ${\cal M}_{N}$ is reflected by
the following relation :
\begin{eqnarray*}
\alpha^{rs}.\alpha^{mn} = q^{(rn-ms)}\alpha^{mn}\alpha^{rs}.
\end{eqnarray*}
We can also equip ${\cal M}_{N}$ with a Lie structure by introducing the
following commutation rule :
\begin{eqnarray*}
[\alpha^{rs} , \alpha^{mn}] = C^{(rs)(mn)}~_{(kl)} \alpha^{kl} 
\end{eqnarray*}
where the structure constants are given by :
\begin{eqnarray*}
C^{(rs)(mn)}~_{(kl)} = (q^{-ms} - q^{-nr})\delta^{r+m}~_{k}~
\delta^{s+n}~_{l}.
\end{eqnarray*}
Let us define a basis $\{ e_{rs} \}$ of Der(${\cal M}_{N}$), i.e.
the Lie algebra of derivations (all are inner) of ${\cal M}_{N}$ as follows:
\begin{eqnarray*}
e_{rs} = Ad_{\alpha^{rs}} = [\alpha^{rs} ,~.~  ]
\end{eqnarray*}
such that
\begin{eqnarray*}
e_{rs}(\alpha^{mn}) = [ \alpha^{rs} , \alpha^{mn}] =
C^{(rs)(mn)}~_{(kl)}\alpha^{kl}
\end{eqnarray*}
and satisfying :
\begin{eqnarray*}
[ e_{rs} , e_{mn} ] = C_{(rs)(mn)}~^{(kl)} e_{kl}.
\end{eqnarray*}

\section{The quantum group ${\cal F}$, its dual ${\cal H}$ and reduction
of ${\cal M}_{N}$}

\subsection{The quantum group ${\cal F}$ and its coaction on ${\cal M}_{N}$ }

Let us construct the matrix quantum group generated by the quantum matrix :
$\left( \begin{array}{cc}
         a & b \\
         c & d
        \end{array}
        \right)$
coacting on the coordinate doublet of the reduced quantum plane by
the following left and right coactions :
\begin{eqnarray*}
\left( \begin{array}{c}
        x' \\
        y'
        \end{array}
        \right) = \delta_{L}\left( \begin{array}{c}
                                     x  \\
                                     y
                                   \end{array}
                                   \right) =
\left( \begin{array}{cc}
          a & b \\
          c & d 
         \end{array}
          \right) \otimes \left( \begin{array}{c}
                                    x \\
                                    y 
                                 \end{array}
                                 \right) = 
\left( \begin{array}{c}
        a \otimes x + b \otimes y \\
        c \otimes x + d \otimes y 
        \end{array}
        \right)
\end{eqnarray*}
\begin{eqnarray*}
\left( x"~y" \right) = \delta_{R} \left( x ~y \right) = 
(x~y) \otimes \left( \begin{array}{cc}
                       a & b \\
                       c & d 
                       \end{array}
                       \right) = 
\left( x \otimes a + y \otimes c~~~x \otimes b + y \otimes d \right).
\end{eqnarray*}
Imposing that the quantities $x'$,$y'$ (and $x"$,$y"$) should satisfy the
same relations as $x$ and $y$, one obtains the following defining relations
of the quantum group Fun($GL_{q}(2)$)
\begin{eqnarray*}
\begin{array}{l}
ab = qba \\
ac = qca \\
ad - da = (q - q^{-1})bc \\
bc = cb \\
bd = qdb \\
cd = qdc 
\end{array}
\end{eqnarray*}
together with :
\begin{eqnarray*}
a^{N} = d^{N} = {\bf 1}~~,~~ c^{N} = b^{N} = 0   .
\end{eqnarray*}
These latter represent an ideal $I$, such that the resulting quantum group is
the quotiented quantum group $\mbox{Fun}(GL_{q}(2))/I$.
The element ${\cal D} = ad - qbc = da - q^{-1}bc$ is central and represents 
the $q$-determinant, and if we set it equal to $1$, we get the
quotiented $\mbox{Fun}(SL_{q}(2))/I$.\\
The algebra defined by $a,b,c,d$ and the above set of relations will be called
${\cal F}$. Using the fact that $a^{N} = {\bf 1}$ and that :
\begin{eqnarray*}
a d  &=&{\bf 1} + q b c
\end{eqnarray*}
we obtain $d = a^{N-1}({\bf 1} + q b c)$, so that d (or $a$) can be
eliminated.\\
The algebra ${\cal F}$ can therefore be linearly generated -as a vector space-
by the elements $a^{\alpha} b^{\beta} c^{\gamma}$ where $\alpha, \beta,
\gamma = 0,1,2,.....,N-1 $. We see that ${\cal F}$ is a finite dimensional
associative algebra, of dimension $N^{3}$. 

\subsection{The quantum algebra ${\cal H}$ and its action on ${\cal M}_{N}$ }

Using the interchange of multiplication and comultiplication by duality, we
define the dual ${\cal H}$ of ${\cal F}$ as a quantum group of same
dimension as ${\cal F}$, generated by ${ H^{\alpha} X_{+}^{\beta}
X_{-}^{\gamma}; \alpha, \beta, \gamma \in z }$,
where $X_{+}, X_{-}, H$ are defined by duality by means of the following
pairing between generators:

\begin{eqnarray*}
\begin{array}{llll}
<H,a> = q~~,&~~<H,b> = 0~~,&~~<H,c> = 0~~,&~~<H,d> = q^{2} \\
<H^{-1},a> = q^{2}~~,&~~<H^{-1},b> = 0~~,&~~<H^{-1},c> = 0~~,&~~<H^{-1},d> =q \\  
<X_{+},a> = 0~~,&~~<X_{+},b> = 1~~,&~~<X_{+},c> = 0~~,&~~<X_{+},d> = 0 \\ 
<X_{-},a> = 0~~,&~~<X_{-},b> = 0~~,&~~<X_{-},c> = 1~~,&~~<X_{-},d> = 0 
\end{array}
\end{eqnarray*}
and the relations:
\begin{eqnarray*}
\begin{array}{rcl}
H^{N} &=& {\bf 1} \\
X_{+}^{N} &=& X_{-}^{N} = 0
\end{array}
\end{eqnarray*}
${\cal H}$ acts on the reduced quantum plane ${\cal M}_{N}$, since its dual
${\cal F}$ coacts on it. There are again two possibilities, left or right, but we
shall use the left action that is generally defined as follows . If we denote
the right coaction of ${\cal F}$ on ${\cal M}_{N}$ as :
\begin{eqnarray*}
\delta_{R}(z) = \sum_{i} z_{i} \otimes u_{i}
\end{eqnarray*}
then :
\begin{eqnarray*}
\begin{array}{rcl}
X_{L}(z) &=& (Id \otimes <X_{L} , . >) \circ \delta_{R}(z) \\
         &=& (Id \otimes <X_{L} , . >)(\sum_{i} z_{i} \otimes u_{i}) \\
         &=& \sum_{i} <X_{L} , u_{i} > z_{i} ,
\end{array}
\end{eqnarray*}
for $z, z_{i} \in {\cal M}, X_{L} \in {\cal H} , u_{i} \in {\cal F}$.\\
It follows that the action of ${\cal H}$ on ${\cal M}$ is given
by the following table :
\begin{eqnarray*}
\begin{tabular}{|c|c|c|c|} \hline
Left     & $H$       & $X_{+}$ & $X_{-}$ \\ \hline
${\bf 1}$& ${\bf 1}$ & $0$     & $0$  \\ \hline
$x$      & $qx$      & $0$     & $y$ \\  \hline
$y$      & $q^{2}y$  & $x$     & $0$  \\ \hline
\end{tabular}
\end{eqnarray*}
For an arbitrary element of ${\cal M}$, one find the following expressions :
\begin{eqnarray*}
\begin{array}{rcl}
H^{L}[x^{r} y^{s}]  &=& q^{(r-s)} x^{r} y^{s}   \\
X_{+}^{L}[x^{r} y^{s}] &=& q^{r}(\frac{1-q^{-2s}}{1-q^{-2}}) x^{r+1} y^{s-1}\\
X_{-}^{L}[x^{r} y^{s}] &=& q^{s}(\frac{1-q^{-2r}}{1-q^{-2}}) x^{r-1} y^{s+1}
\end{array}
\end{eqnarray*}
with $r, s = 0,1,2,.......,N-1 $.

\subsection{The reduction of the algebra ${\cal M}_{N}$ into
indecomposable representation of ${\cal H}$ }

The generator $H$ always acts as an automorphism, for this reason, in order
to study the invariant subspaces of ${\cal M}_{N}$ under the left action of
${\cal H}$, we have only to consider the action of $X_{+}$ and $X_{-}$.\\
Forgetting numerical factors, the action of $X_{+}$ and
$X_{-}$ on a given element of ${\cal M}$ can be written as follows :
\begin{eqnarray*}
x^{r+1} y^{s-1} \rightleftharpoons
x^{r} y^{s} \rightleftharpoons
x^{r-1} y^{s+1}
\end{eqnarray*}
where $X_{-}$ takes us from the left to the right and $X_{+}$ from the right
to the left.\\
We verify that under the left action of ${\cal H}$ the algebra of $N{\times}N$
matrices can be decomposed into a direct sum of $N$ subspaces of dimension
$N$, according to :
\begin{eqnarray*}
\begin{array}{rcl}
N_{N}~~ &=& \{x^{N-1}, x^{N-2}y, x^{N-3}y^{2}, x^{N-4}y^{3}, ......., xy^{N-2}, y^{N-1} \} \\
N_{N-1} &=& \{x^{N-2}, x^{N-3}y, x^{N-4}y^{2}, ......., xy^{N-3}, y^{N-2}, x^{N-1}y^{N-1} \} \\
N_{N-2} &=& \{x^{N-3}, x^{N-4}y, x^{N-5}y^{2}, ......., xy^{N-4}, y^{N-3}, x^{N-1}y^{N-2}, x^{N-2}y^{N-1} \} \\
N_{N-3} &=& \{x^{N-4}, x^{N-5}y, x^{N-6}y^{2}, ......., xy^{N-5}, y^{N-4}, x^{N-1}y^{N-3}, x^{N-2}y^{N-2}, x^{N-3}y^{N-1}\} \\
N_{N-4} &=& \{x^{N-5}, x^{N-6}y, x^{N-7}y^{2}, ......., xy^{N-6}, y^{N-5}, x^{N-1}y^{N-4}, x^{N-2}y^{N-3}, x^{N-3}y^{N-2}, \\
        &~& x^{N-4}y^{N-1} \} \\
.  \\
.  \\
.  \\
N_{2}~~ &=& \{x, y, x^{N-1}y^{2}, x^{N-2}y^{3}, ......., x^{3}y^{N-2}, x^{2}y^{N-1} \} \\
N_{1}~~ &=& \{{\b 1}, x^{N-1}y, x^{N-2}y^{2}, x^{N-3}y^{3}, ......., x^{2}y^{N-2}, xy^{N-1} \} \\
\end{array}
\end{eqnarray*}
such that :
\begin{eqnarray*}
{\cal M}_{N} = N_{N} {\oplus} N_{N-1} {\oplus} ..............{\oplus} N_{2} {\oplus} N_{1}
\end{eqnarray*}

\section{The $Z_{N}$--graded differential geometry of $Z_{N}$}

First, let us recall that it is possible to construct a $Z_{2}$--graded
noncommutative differential geometry of ${\cal M}_{N}$ based on derivations, by
introducing a set of 1--forms $\theta^{kl}$ defined by the following
duality relation, \cite{2} :
\begin{eqnarray*}
\theta^{kl} \left( e_{mn} \right) = \delta^{kl}~_{mn} = \delta^{k}~_{m}
\delta^{l}~_{n}.
\end{eqnarray*}
Then, using the $Z_{2}$--graded differential $d$ (and the wedge product), one
easily describe the $Z_{2}$--graded noncommutative differential complex
$\left( \Omega_{Der}({\cal M}_{N}) ; d \right)$.\\
Our main aim in this work is precisely to show that ${\cal M}_{N}$ itself,
equiped with some well--defined differential $d$ satisfying $d^{N} = 0$,
can be viewed as a $Z_{N}$--graded differential complex of the cyclic group
$Z_{N}$.\\
For this purpose, let us define a $Z_{N}$--{\em grading} on ${\cal M}_{N}$
such that :
\begin{eqnarray*}
|\alpha^{rs}| = \mbox{grading}(\alpha^{rs}) = r+s ~~~\mbox{mod(N)}.
\end{eqnarray*}
This means that a $Z_{N}$--grading equal to 1 is attributed to the
fundamental objects ${\bf 1}$, $x$ and $y$, and then the above decomposition
of ${\cal M}_{N}$ is naturally equiped with the following $Z_{N}$--grading :
\begin{eqnarray*}
\begin{array}{l}
N_{1}~~\longrightarrow~~0 \\
N_{2}~~\longrightarrow~~1 \\
N_{3}~~\longrightarrow~~2 \\
. \\
. \\
N_{N-2}~~\longrightarrow~~N-3 \\
N_{N-1}~~\longrightarrow~~N-2 \\
N_{N}~~~\longrightarrow~~N-1 \\
\end{array}
\end{eqnarray*}
Consider the cyclic group of order $N$,
$Z_{N} = \{ {\bf 1}, y, y^{2}, y^{3},..........., y^{N-1}\} $.Therefore, the
algebra $C^{\infty}(Z_{N})$ of complex functions on $Z_{N}$ can be realized
as the algebra ${\cal M}_{N}^{diag} \subset {\cal M}_{N}$ of diagonal
complex $N{\times}N$ matrices.\\
Starting from $C^{\infty}(Z_{N}) {\equiv}{\Omega}^{0}(Z_{N}) = Z_{N}$, we
can build the space of 1--forms $\Omega^{1}(Z_{N})$ by introducing a differential
$d_{x} : \Omega^{0} \longrightarrow \Omega^{1}$ associated to $x$ and
defined by :
\begin{eqnarray*}
d_{x}(y^{m}) = [x , y^{m} ] = (1 - q^{-m}) xy^{m}.
\end{eqnarray*}
This means that the sub--space
$\Omega^{1} =x\Omega^{0} =  \left\{ x , xy , .... ,xy^{N-1} \right\}$
constitutes the space of 1--forms. \\
This differential can be naturally extended to all other sub--spaces of
${\cal M}_{N}$ such that :
\begin{eqnarray*}
d_{x} : \Omega^{k} \longrightarrow \Omega^{k+1}
\end{eqnarray*}
\begin{eqnarray}
d_{x}(\alpha^{rs}) = x\alpha^{rs} - q^{r} \alpha^{rs}x = [ x , \alpha^{rs} ]_{q} =
(1 - q^{r-s})\alpha^{(r+1),s}  \label{a}
\end{eqnarray}
where the sub--space of $k$--forms is defined by : 
\begin{eqnarray*}
\Omega^{k} = x^{k} \Omega^{0} =\left\{ x^{k}, x^{k}y, ...., x^{k}y^{N-1} \right\}
\end{eqnarray*}
for $k=0,1,....,N-1$. It is easy to see that the {\em degree} of the
differential forms is given by :
\begin{eqnarray*}
\mbox{degree}(\alpha^{rs}) = r ~~~~\mbox{mod(N)}
\end{eqnarray*}
and that the wedge product between two arbitrary forms is nothing else than
the usual matrix multiplication.\\
Then, the $Z_{N}$--graded differential complex  $\left({\Omega}(Z_{N}),
d\right)$, with $d^{N}=0$, is completely built with :
\begin{eqnarray*}
{\Omega}(Z_{N}) = {\Omega}^{0} {\oplus} {\Omega}^{1} {\oplus} {\Omega}^{2}
{\oplus} .........{\oplus} {\Omega}^{N-2} {\oplus} {\Omega}^{N-1}~
~{\approx}M_{N}(C)
\end{eqnarray*}
Moreover, one can easily verify that the differential $d$ satisfy a
$q$-deformed Leibniz rule :
\begin{eqnarray*}
d_{x}(\alpha^{rs} \alpha^{mn}) = \left( d_{x}(\alpha^{rs}) \right)
\alpha^{mn}  + q^{r} \alpha^{rs} \left( d_{x}(\alpha^{mn})
\right)
\end{eqnarray*}
and that effectively one has $d^{N}=0$.

\section{EXAMPLE : The cyclic group $Z_{3}$}

Let us now consider the case of $Z_{3} = \{{\bf 1},y,y^{2}\}$, with :
\begin{eqnarray*}
{\bf 1} = \left( \begin{array}{ccc}
          1 & 0 & 0 \\
          0 & 1 & 0 \\
          0 & 0 & 1
          \end{array}
       \right) = y^{3}~~,~~
y = \left( \begin{array}{ccc}
           1 & 0 & 0 \\
           0 & q & 0 \\
           0 & 0 & q^{2}
          \end{array}
    \right)~~,~~
y^{2} = \left( \begin{array}{ccc}
                1 & 0     & 0 \\
                0 & q^{2} & 0 \\
                0 & 0     & q
               \end{array}
        \right).
\end{eqnarray*}
\begin{eqnarray*}
1 + q + q^{2} = 0 ~~~~,~~~q = q^{-2}~~~,~~~q^{2} = q^{-1}~~,q^{3} = 1.
\end{eqnarray*}
The algebra $C^{\infty}(Z_{3})$ of complex functions on $Z_{3}$ is then
identified with the sub--algebra ${\cal M}_{3}^{diag} \subset {\cal M}_{3}$
of diagonal complex $3 \times 3$ matrices, where ${\cal M}_{3}$ is
generated by :
\begin{eqnarray*}
\{{\bf 1}, x, y, xy, x^{2}, y^{2}, x^{2}y, xy^{2}, x^{2}y^{2} \} ,
\end{eqnarray*}
with
\begin{eqnarray*}
x = \left( \begin{array}{ccc}
          0 & 1 & 0 \\
          0 & 0 & 1 \\
          1 & 0 & 0
          \end{array}
          \right).
\end{eqnarray*}
If we attribute a $Z_{3}$--grading $1$ to ${\bf 1}$, $x$ and $y$, then one has :
\begin{eqnarray*}
\begin{array}{rcl}
\{{\bf 1}, xy^{2}, x^{2}y\} \longrightarrow 0  \\
\{x, y, x^{2}y^{2}\} \longrightarrow1   \\
\{x^{2}, y^{2}, xy\} \longrightarrow2
\end{array}
\end{eqnarray*}
From the sub--space $\Omega^{0} = Z_{3}$ of 0--forms, we build the two
other subspaces of 1-- and 2--forms respectively :
\begin{eqnarray*}
\begin{array}{rcl}
\Omega^{1} &=& x\Omega^{0} = \{x, xy, xy^{2} \}\\
\Omega^{2} &=& x^{2}\Omega^{0} = \{x^{2}, x^{2}y, x^{2}y^{2} \} . 
\end{array}
\end{eqnarray*}
by using the differential $d_{x} : \Omega^{k} \longleftarrow \Omega^{k+1}$
defined by (\ref{a}), i.e. :
\begin{eqnarray}
\begin{array}{l}
d_{x}({\bf 1 }) = 0   \\
d_{x}(y)       = (1-q^{2})xy   \\
d_{x}(y^{2})   = (1-q)xy^{2}  \\
d_{x}(x)       = (1-q)x^{2}    \\
d_{x}(xy)      =  0  \\
d_{x}(xy^{2})  = (1-q^{2}) x^{2}y^{2}  \\
d_{x}(x^{2})   = (1-q^{2}){\bf 1}  \\
d_{x}(x^{2}y)  = (1-q)y   \\
d_{x}(x^{2}y^{2}) = 0  \\
\end{array} \label{b}
\end{eqnarray}
Then, the $Z_{3}$--graded differential algebra ${\Omega}(Z_{3})$ is given by :
\begin{eqnarray*}
{\Omega}(Z_{3}) = {\Omega}^{0} {\oplus} {\Omega}^{1} {\oplus} {\Omega}^{2}~
~{\approx} M_{3}(C)
\end{eqnarray*}
with :
\begin{eqnarray*}
{\Omega}^{k} = x^{k}Z_{3} ;~~~~~ k = 0,1,2 .
\end{eqnarray*}
Finally, using the relations (\ref{b}), we can easily verify that for
arbitrary $\omega_{p} \in \Omega^{p}(Z_{N})$
and $\omega_{q} \in \Omega^{q}(Z_{N})$ one has :
\begin{eqnarray*}
d_{x}(\omega_{p}\omega_{q}) = (d_{x}\omega_{p})\omega_{q} +
q^{p}\omega_{p}(d_{x}\omega_{q})
\end{eqnarray*}
and 
\begin{eqnarray*}
\begin{array}{rcl}
d^{3}_{x}(\omega_{p}) &=& [x,[x,[x,\omega]_{q}]_{q}]_{q}          \\
                      &=& [x,[x,(x\omega - q^{k}\omega x)]_{q}]_{q}\\
                      &=& ............\\
                      &=& q^{k}(1+q+q^{2})[.....] + x^{3}\omega - \omega x^{3} \\
                      &=& 0.
\end{array}
\end{eqnarray*}

\section{Conclusion}

In the last decade, noncommutative differential geometry became a very
important research topic in Mathematical Physics. In this context, the role of the
$C^{*}$--algebra of smooth complex functions on a ordinary manifold is played
by an abstract {\em associative} not neces\-sarily commutative $C^{*}$--algebra
as analog of functions on noncommutative manifolds.\\
In order to define gauge theories on these noncommutative spaces,
we need to define noncommutative differential calculus on them. In fact,
several particle Physics models have been constructed on noncommutative
spaces, for instance, on product spaces like $C^{\infty}(M) \otimes M_{N}(C)$,
${\cal M}_{4} \times Z_{N}$, etc... , \cite{2}, \cite{5}.\\
In another hand, the matrix algebras ${\cal M}_{N}$ are very often used in
various fields of Physics. Moreover, its differential geometry
is the most simple example of noncommutative differential geometry.
In \cite{3}, the Wess--Zumino complex of ${\cal M}_{N}$ is constructed. \\
Nevertheless, following the Dubois-Violette's approach, \cite{2}, we show how
to construct the biggest sub--algebra of the noncommutative
universal differential algebra of these matrix algebras, and present its
decomposition into irreducible components by determining the eigenvalue
equations of the associated Laplace--Beltrami operator, with a special
interest to the case of $M_{3}(C)$, \cite{6}. \\
Actually, it seems very interesting to study the $Z_{N}$--graded 
differential geometry of some noncommutative spaces. We plan to treat this
subject in a future paper in order to \-describe gauge theories on such spaces.\\
\vskip2truecm
\section*{Acknowledgments }

The authors would like to acknowledge Abdus Salam International Centre for
Theoretical Physics where this work was realized under the Associateship
scheme. They also would like to thank the Arab fund for financial support.


\newpage

\begin{thebibliography}{99}
\bibitem{1}{\bf A.Connes}~:~ Publ. IHES 62(1986)257;
\bibitem{2}{\bf M.Dubois-Violette and al}~:~ J.Math.Phys. 31(1990)316;  \\
see also the review : \\
{\bf A.E.F. Djemai}, Int.J.Theor.Phy. 34(1995)801 .
\bibitem{3}{\bf R. Coquereaux and al}~:~ "{\em Finite dimensional quantum
group covariant differential calculus on a complex matrix algebra }", math.QA/
9804021,\\
{\bf R. Coquereaux and G.E. Schieber}~:~"{\em Action of a finite quantum
group on the algebra of complex $N \times N$ matrices}", math-ph/9807016
\bibitem{4}{\bf M. Dubois-Violette}~:~ "{\em Generalized differential geometry
with $d^{N} = 0$ and the $q$--differential calculus}" LPTHE--Orsay 96/75,
Czech J. Phys. 46(1997)1227,\\
{\bf M. Dubois-Violette and R. Kerner}~:~ "{\em Universal $Z_{N}$--graded
differential calculus}" LPTHE--Orsay (1996),
\bibitem{5}{\bf R. Coquereaux}~:~J. Geom. Phys. 11(1993)307-324,\\
{\bf A. Sitarz}~:~J. Geom. Phys. 15(1995)123-136,\\
{\bf Y. Okumura}~:~Prog. Theor. Phys. 96(1996)1021-1036,
\bibitem{6}{\bf A.E.F. Djemai and H. Smail}:"{\em The Non--commutative
Differential Geometry of the Matrix Algebra $M_{N}(C)$ }, to be submitted
for publication.
\end{thebibliography}
\end{document}